# White matter deficits underlie the loss of consciousness level and predict recovery outcome in disorders of consciousness

Xuehai Wu[a], Jiaying Zhang[b], Zaixu Cui[b], Weijun Tang[c], Chunhong Shao[d], Jin Hu[a], Jianhong Zhu[a], Liangfu Zhou[a], Yao Zhao[a], Lu Lu[e], Gang Chen[f], Georg Northoff[g], Gaolang Gong[b*], Ying Mao[a*], Yong He[b]

[a] Neurosurgical Department, Shanghai Huashan Hospital, Fudan University, Shanghai 200040, China

[b] State Key Laboratory of Cognitive Neuroscience and Learning & IDG/McGovern Institute for Brain Research, Beijing Normal University, Beijing, 100875 China

[c] Radiological Department, Shanghai Huashan Hospital, Fudan University, Shanghai 200040, China

[d] Psychiatry Department, Shanghai Huashan Hospital, Fudan University, Shanghai 200040, China

[e] Huajia Hospital, Shanghai 200438, China

[f] Scientific and Statistical Computing Core, National Institute of Mental Health, National Institutes of Health, Department of Health and Human Services, USA

[g] Institute of Mental Health Research, University of Ottawa, Carling Avenue 1145, Ottawa, ON K1Z 7K4, Canada

Xuehai Wu and Jiaying Zhang contributed equally to this work.

[*] Corresponding authors:

Ying Mao

Neurosurgical Department, Shanghai Huashan Hospital,

Fudan University, Shanghai, China

Phone: +8613801769152

Email: maoying@fudan.edu.cn

Gaolang Gong

State Key Laboratory of Cognitive Neuroscience and Learning & IDG/McGovern

Institute for Brain Research,

Beijing Normal University, Beijing, China

Phone: +8601058802023

Email: gaolang.gong@bnu.edu.cn

**Abstract**

This study aimed to identify white matter (WM) deficits underlying the loss of consciousness in disorder of consciousness (DOC) patients using Diffusion Tensor Imaging (DTI) and to demonstrate the potential value of DTI parameters in predicting recovery outcomes of DOC patients. With 30 DOC patients (8 comatose, 8 unresponsive wakefulness syndrome/vegetative state, and 14 minimal conscious state) and 25 patient controls, we performed group comparison of DTI parameters across 48 core WM regions of interest (ROIs) using Analysis of Covariance. Compared with controls, DOC patients had decreased Fractional anisotropy (FA) and increased diffusivities in widespread WM area. The corresponding DTI parameters of those WM deficits in DOC patients significantly correlated with the consciousness level evaluated by Coma Recovery Scale – Revised (CRS-R) and Glasgow Coma Scale (GCS). As for predicting the recovery outcomes (i.e., regaining consciousness or not, grouped by their Glasgow Outcome Scale >2 or not) at 3 months post scan, radial diffusivity of left superior cerebellar peduncle and FA of right sagittal stratum reached an accuracy of ~87.5% and ~75% respectively. Our findings showed multiple WM deficits underlying the loss of consciousness level, and demonstrated the potential value of these WM areas in predicting the recovery outcomes of DOC patients who have lost awareness of the environment and themselves.

# 1. Introduction

With great advances in intensive care, more and more patients survive after severe brain injuries such as traumatic brain injury (TBI), spontaneous intracerebral hemorrhage (ICH), and ischaemic-hypoxic injury (IHI), resulting in a large population of patients with disorders of consciousness (DOC). DOC patients exhibit varied levels of consciousness (from low to high: coma (COMA), unresponsive wakefulness syndrome/vegetative state (UWS/VS), and minimally conscious state (MCS)), which provides a unique opportunity to study the structural substrates that support the maintenance of a normal consciousness level. The pioneer functional magnetic resonance imaging (fMRI) study revealed that UWS/VS patients were capable of following a command and performing an imaginary task(Owen et al., 2006). This revolutionized our understanding of the abnormal state of consciousness, but greatly challenged the diagnosis criteria for DOC patients. Sensitive and robust imaging biomarkers will greatly benefit and aid the clinical assessments. Improvements of clinical assessments are key to making reasonable choices of therapy plans and legal or ethical decisions for DOC patients(Monti et al., 2010).

The altered consciousness level of DOC patients is associated with functional connectivity deficiencies across brain regions in previous studies. For example, using PET, those functional connections between intralaminar thalamic nuclei, the anterior cingulate and prefrontal cortices were shown to be disrupted in VS patients(Laureys et al., 2000). Given these findings, the loss of consciousness has been described as a "functional disconnection syndrome"(Laureys et al., 2004; Schiff, 2005). Additionally, recent functional MRI studies observed the association of levels of consciousness with functional connectivity deficiencies between the default mode network (DMN) and the thalamus(Boly et al., 2008; Boly et al., 2009; Vanhaudenhuyse et al., 2010). Whole-brain functional network analysis exhibited abnormal connectivity pattern in DOC patients (Qin et al., 2015). Moreover, those functional connections could predict the consciousness level and recovery outcome of DOC patients (Wu et al., 2015). These results further support the idea that human consciousness is an outcome of functional integration among distributed brain regions rather than local functional activities.

Although altered functional connectivity has been investigated a lot, the structural neural substrate of the consciousness loss remains poorly understood. White matter (WM) harbors the main structural connections between different brain regions; besides, clinically, diffused axonal injury patients with the loss of consciousness usually have damaged white matter detectible in Computer Tomography/MRI. Therefore, WM impairments in DOC patients play a critical role in the loss of consciousness. Using diffusion MRI, a pioneer case study of one MCS patient revealed an increase of WM anisotropy, which was related to the recovery of expressive language(Voss et al., 2006). Several cross-sectional studies have revealed that the diffusion parameters of the entire WM and WM pathways connecting the thalamus and DMN brain regions varied depending on the consciousness levels of DOC patients(Fernandez-Espejo et al., 2011; Fernandez-Espejo et al., 2012; Lant et al., 2016; Newcombe et al., 2010). But there is still not a complete understanding of WM substrate underlying the loss of consciousness level. However, these WM studies were limited by (1) recruiting only a small number of subjects (making clinical relevance impossible) or a relatively large sample but from different scanning centers (introducing a new confounding factor of scanning center); (2) mainly covering UWS/VS and MCS patients without including COMA patients; (3) the absence of lesion patients with full consciousness; and (4) a focus on a limited number of pre-defined WM tracts/regions of interest (ROIs) or the WM as a whole. Furthermore, it is clinically unknown whether the diffusion parameters of WM deficits can be used to predict recovery outcomes (regaining consciousness or not) for those patients who have been clinically diagnosed as "unaware of the surrounding environment and themselves". This issue is of great clinical importance for therapeutic planning and decision-making.

Therefore, our study aimed to 1) identify the impaired WM regions underlying the loss of consciousness in DOC patients; and 2) assess the prognostic value of WM diffusion parameters for predicting the recovery outcomes (i.e., outcome-positive and outcome-negative) of DOC patients who have lost awareness of the environment and themselves. Specifically, we included a single-center dataset from a large cohort of 30 DOC patients covering the entire spectrum of consciousness levels (including COMA, UWS/VS, and MCS) as well as 25 patient controls (PCs) from a single center. Given previous findings, we hypothesized that distributed WM injury, rather than a local WM injury, contributes to the abnormal levels of consciousness in these patients. To test this hypothesis, an automated atlas-based searching for the abnormal WM regions using DTI parameters was applied across the whole brain. We then evaluated the relationship of DTI metrics and the clinical measures of consciousness levels across DOC patients. Finally, we performed ROC analysis to predict the clinical recovery outcomes of COMA and UWS/VS patients at 3 months after the MRI scan (regained consciousness or not) using DTI parameters of white matter deficits in DOC patients.

## 2. Materials and methods

### 2.1. Participants

We recruited 91 patients with acquired brain injuries in the Huashan Hospital of Fudan University in Shanghai. The etiology of these patients was either TBI (77 subjects) or non-TBI—spontaneous intracerebral hemorrhage or ischaemic-hypoxic injury (14 subjects). All of the patients participated in the MRI scan during their sub-acute or chronic stage. The majority of the DOC patients were followed up in the rehabilitation hospital, and the others were followed up when be readmitted for ventricular peritoneal shunt with hydrocephalus or/and cranioplasty. Informed consent was obtained from each patient or his/her family. Ethical approval for this study was granted by the Ethics Committee of Huashan Hospital.

The patients were clinically diagnosed as COMA, UWS/VS, MCS, or Patient Controls (PC) on the day of their first MRI scan. The COMA patients were characterized by the absence of arousal responses and awareness(Plum and Posner, 1966). The UWS/VS patients were in a state of arousal, including a sleep/wake cycle, but were unaware of themselves or their global environment(The Multi-Society Task Force on PVS, 1994). The MCS patients retained a low level of consciousness but exhibited inconsistent and non-reflexive behavior(Giacino et al., 2002). The PCs in our study were characterized by a) a conscious state and communicability; b) a brain injury confirmed by MRI and computed tomography scan; and c) the absence of locked-in syndrome. The consciousness level of each patient was quantified using two standardized scales [the Glasgow Coma Scale (GCS)(Teasdale and Jennett, 1974) and the Coma Recovery Scale-Revised (CRS-R)(Giacino et al., 2004)]. Both scales were translated into Chinese by 3 authors (P.Q., Z.H., and Xu. Wu.). An experienced neurosurgeon, Xu. Wu., assessed both scales and was blind to data analysis.

After manually checking the quality of the fractional anisotropy (FA) images after normalization (exclusion examples are shown in Supplementary Fig. 1), we included 55 patients (8 COMA, 8 UWS/VS, 14 MCS, and 25 PC) in the cross-sectional analysis to detect consciousness-related WM regions. The detailed demographic and clinical characteristics are shown in Table 1 and Supplementary Fig. 2. Moreover, for those patients who had completely lost their awareness of the environment and themselves at their first MRI scan (8 COMA and 8 UWS/VS patients), we divided them into outcome-positive and outcome-negative groups according to their clinical outcomes (regained awareness or not) at least 3 month after MRI scan. Specifically, the Glasgow Outcome Scale (GOS)(Jennett and Bond, 1975) was used to quantify clinical outcomes 3 months after the first MRI scan. The patients with a GOS score of less than 3 were deemed as not having regained consciousness and were referred to as outcome-negative. Patients in the other group, with a GOS score equal to or above 3, were considered to be awake and were therefore referred to as outcome-positive. The corresponding demographic and clinical characteristics of the outcome-negative and outcome-positive groups are shown in Table 2.

*2.2. Diffusion Weighted Images Acquisition*

All MRI images were acquired on the 3T SIEMENS MRI scanner in Shanghai Huashan Hospital. We used a single-shot echo planar imaging-based sequence, and the acquisition protocol consisted of 12 non-linear diffusion-weighted directions with b = 1000 s/mm$^2$ and one additional image without weighted diffusion (i.e., b = 0 s/mm$^2$). The scanning parameters were set as follows: 3.5 mm slice thickness, no gap between slices, 38 slices covering the whole brain, echo time = 82 ms, repetition time = 8400 ms, acquisition matrix = 128×128, non-interpolated voxel size = 1.8×1.8×3.5 mm$^3$, flip angle = 90°, and field of view = 230×230 mm$^2$.

*2.3. DTI post-processing*

The diffusion-weighted images were processed with a PANDA pipeline toolbox(Cui et al., 2013) called the FMRIB Software Library (FSL, version 4.1.9)(Jenkinson et al., 2012) for skull stripping, eddy-current correction, tensor fitting and the calculation of diffusion tensor parameters as well as for image normalization. Common diffusion tensor parameters—FA, axial diffusivity (AD), and radial diffusivity (RD)—were chosen for subsequent analysis. Specifically, FA is the fraction of anisotropic diffusion, and a breakdown in WM integrity typically results in a lower FA(Basser and Pierpaoli, 1996). AD and RD are thought to be selectively sensitive to specific microstructural changes. An increased AD reflects axonal damage or loss, whereas an increased RD reflects demyelination(Song et al., 2003). Another common diffusion parameter, mean diffusivity (MD), was not included in the present study because it is linearly dependent on AD and RD. After nonlinearly normalizing the FA, AD, and RD maps to the FMRIB58_FA template, we extracted the FA, AD, and RD of the 48 WM ROIs defined in the standard space of the ICBM-DTI-81 white matter atlas(Mori et al., 2008).

In the present study, we performed the analysis of WM parameters at the regional level, primarily for two reasons. First, ROI-based analysis exhibits a higher tolerance for inaccurate normalization compared with voxel-based analysis(Faria et al., 2010). Second, it is biologically plausible to assume diffuse WM impairment following traumatic or ischaemic/hypoxic injury rather than a very local and concentrated impairment. Additionally, because outer brain tissues may be severely damaged in TBI patients, we defined the WM regions according to the JHU "core white matter" atlas(Mori et al., 2008). The JHU "core white matter" atlas consists of 48 WM regions in MNI152 space that are located relatively deep inside the brain and are therefore less deformed in DOC patients (Supplementary Fig. 3). In all, region-based analysis could provide adequate spatial specificity and improve the statistical power for our study.

*2.4. Statistical analysis*

All statistical analyses were performed with IBM SPSS 20 statistics software. We first compared the demographic and clinical data across the four patient groups. One-way analysis of variance (ANOVA) and chi-square tests were performed on the continuous variables (age and days post-ictus) and categorical variables (etiology and gender), respectively.

To explore the association between WM deficits and consciousness level, we first performed a one-way analysis of covariance (ANCOVA) on the mean FA for each of the 48 WM regions. Specifically, group was a between-subjects factor of interest, and age, gender, and days post-ictus were included as covariates. The Bonferroni method was applied to correct for multiple comparisons across the WM regions, and a corrected $p < 0.05$ was considered significant. To identify whether the FA changes were attributed to AD and/or RD changes, we applied the same ANCOVA model to the mean AD and RD of the WM deficits detected by FA ANCOVA.

Among the WM deficits detected by FA ANCOVA, Pearson correlations between the mean FA and clinical scores of consciousness level (GCS and CRS-R) were tested across all of the patients after controlling for age, gender, and days post-ictus. Similar correlation analyses were applied to the AD and RD of the consciousness-related WM deficits showing significant group effects in AD or RD. Notably, a substantial proportion of the patients (mainly the PC and UWS/VS patients) received ceiling-level scores (GCS score equal to 15 or CRS-R score equal to 23), which might bias the correlation results. To control for this ceiling effect (the correlation might be driven by the high GCS and CRS-R scores at the upper end), we also performed the correlations after excluding the patients with ceiling-level scores.

To this point, we had explored whether the WM deficits detected by FA ANCOVA could predict the degree of consciousness at different levels of the consciousness spectrum. However, the question remained as to whether DTI metrics of these WM deficits could be used to predict clinical recovery of consciousness. Therefore, we further evaluated the prognostic value of the diffusion parameters of the WM deficits using receiver operating characteristic (ROC) curves. We performed the prediction analysis within white matter ROIs with group differences in DTI parameters (FA, AD, and RD) between the outcome-positive (GOS greater than or equal to 3) and outcome-negative patients (GOS less than 3). The comparisons between the two groups with different outcomes were restricted to the WM deficits detected by FA ANCOVA. Because the level of consciousness may relate to subsequent clinical outcomes, we also included the clinical scores (GCS and CRS-R), age, gender, and days post-ictus as covariates.

3. Results

*3.1. Clinical samples*

There were no differences among the four patient groups (i.e., COMA, UWS/VS, MCS, and PC) in age, gender, or etiology distribution. The duration of illness (days post-ictus) differed ($p = 0.009$). In the patients of the outcome-positive and outcome-negative groups who lost awareness, no differences were found in gender, etiology distribution, or days post-ictus, but there was a trend for the age distribution ($p = 0.052$). The mean age (38.9) of the outcome-positive group was younger than that (49.8) of the outcome-negative group.

*3.2. Group differences in diffusion parameters across DOC patients*

The ANCOVA of FA revealed significant group effects in 14 WM regions (Table 1 and Fig. 1). For most of the 14 WM regions, post hoc comparisons indicated that FA decreased as the consciousness level declined. However, the significant group effect might be mainly driven by the differences between the COMA and PC patients (e.g., the body of the corpus callosum).

Among the 14 WM regions detected by FA ANCOVA, there were significant group differences in AD in only three WM regions (Table 1 and Supplementary Fig. 4), whereas RD significantly differed in 11 WM regions (Table 1 and Supplementary Fig. 5). As the level of consciousness decreased, the mean RD in the 11 WM regions increased.

*3.3. Correlations between diffusion parameters and GCS/CRS-R clinical scores*

With all the patients, significant correlations ($p < 0.05$) were found between the GCS/CRS-R scores and the mean FA of the 14 significant WM regions (as described above) (Figs. 2 and 3). Even after the patients with GCS scores equal to 15 were excluded, 12 WM regions were still correlated with GCS score. In contrast, no correlations were found between the mean FA of these regions and CRS-R score after the patients with a CRS-R score equal to 23 were excluded.

For the three WM regions with AD differences, the mean AD was correlated with the GCS and CRS-R scores across all subjects, but no correlation existed after the patients with ceiling-level scores (GCS or CRS-R) were excluded (Supplementary Figs. 6 and 7). In the 11 WM deficits with RD differences, the mean RD correlated with GCS and CRS-R score. After the exclusion of the patients with maximum GCS and CRS-R scores, the mean RD of eight WM regions (i.e., the body of the corpus callosum, splenium of the corpus callosum, left cingulum, fornix (column and body), left stria terminalis, right internal capsule (retrolenticular portion), right sagittal stratum, and left uncinate fasciculus) still correlated with the GCS score. However, CRS-R score did not significantly correlate with the mean RD of any region (Supplementary Figs. 8 and 9).

Taken together, our data confirm the relevance of WM deficits for predicting the level of consciousness from the lowest end (COMA) to the medium (UWS/VS), higher (MCS), and highest end (PC) of the spectrum of consciousness. Even after controlling for the ceiling effect, the majority of the correlations remained the same; this finding suggests that these correlations were driven by differences across the spectrum of consciousness, from COMA to PC.

### 3.4. Prediction analysis for the outcome-positive and outcome-negative patients

To demonstrate the clinical relevance of our findings, we performed classification analyses of the outcome-positive and outcome-negative patients. The comparisons indicated that the mean FA of the right sagittal stratum and the mean RD of the left superior cerebellar peduncle differed significantly between these two groups with different clinical outcomes at 3 months after scan (Fig. 4). According to the ROC analysis, the prediction accuracy of the mean RD of the left superior cerebellar peduncle and the mean FA of the right sagittal stratum reached 87.5% and 75%, respectively. The corresponding sensitivity and specificity for the mean RD of the left superior cerebellar peduncle were 100% and 75%, respectively, and the sensitivity and specificity for the mean FA of the right sagittal stratum were 50% and 100%, respectively.

## 4. Discussion

Our study for the first time demonstrated widespread WM deficits underlying the diverse consciousness levels of DOC patients. And of those WM deficits, DTI metrics were linearly correlated with the clinical severity of their consciousness level—assessed by GCS and CRS-R scores. Furthermore, our prediction results indicated that the accuracy of distinguishing different clinical outcomes at three months after the MRI scan using DTI parameters of white matter deficits

### *4.1. White matter deficits in DOC patients*

By searching the entire "core white matter", we identified 14 WM regions in which the FA differed across levels of consciousness using ANCOVA, supporting our hypothesis that a widespread structural substrate of the loss of consciousness were in DOC patients. Specifically, most of these WM deficits had increased RD as the consciousness level declined across the DOC subgroups, suggesting a process of axonal demyelination in the DOC patients(Song et al., 2003).

Consistent with the previous DTI study in VS and MCS patients (Fernandez-Espejo et al., 2012), we identified right internal capsule (retrolenticular portion) and the cingulum with abnormal DTI parameters, which may serve as an anatomical substrate for the thalamo-cortical functional connectivity deficiencies in DOC (Boly et al., 2008; Boly et al., 2009; Schnakers, 2012). In addition to these tracts, we also found the splenium and body of the corpus callosum exhibited DTI parameter changes in DOC patients, indicating that the loss of consciousness that occurs in DOC may be related to the disruption of functional interactions between the two hemispheres(Newcombe et al., 2010).

### *4.2 Roles of the white matter around the brainstem*

More importantly, our results provided further evidence of the importance of the brainstem in maintaining normal consciousness. Intriguingly, our results showed that five WM regions around the brainstem, including the middle cerebellar peduncle, right corticospinal tract, right superior cerebellar peduncle, left superior cerebellar peduncle, and right cerebral peduncle, were related to the level of consciousness. The brainstem is crucial for regulation of the sleep cycle and the maintenance of consciousness(Moruzzi and Magoun, 1949; Northoff, 2014;

Parvizi and Damasio, 2003; Starzl et al., 1951). The white matter around the brainstem serves as the main pathway that links the peripheral organs and cerebellum with the cerebrum. Our findings suggested the disruption of information flow not only within the brain but also between the brain and the rest of the body. For example, the normal impulses from peripheral nervous system about the surrounding environment might be disrupted in DOC patients. This finding suggests a possible role of these non-cerebral structures in the normal expression of consciousness, which deserves further attention in future studies.

### *4.3. Clinical relevance*

With more patients surviving after severe brain injuries, it has become an important issue to accurately assess the level of consciousness in DOC patients. However, currently, the clinical diagnosis of DOC patients is relatively subjective and greatly depends on clinical experience(Coleman et al., 2009); as a result, the misdiagnosis rate is high (up to 40%)(Schnakers et al., 2009). The correlations between the diffusion parameters of the WM deficits and the level of consciousness suggest that these imaging parameters have the potential to aid the clinical diagnosis of this DOC patient population. In addition, it remains challenging to accurately predict whether severe DOC patients such as COMA and UWS/VS patients could regain awareness. Our ROC analysis demonstrated the potential value of DTI parameters—FA in the right sagittal stratum and RD in the left superior cerebellar peduncle—in predicting the patients' clinical outcomes at three months after the MRI scan. This finding suggests that the diffusion parameters of specific consciousness-related WM deficits might be sensitive imaging biomarkers to use in predicting the functional recovery of DOC patients.

### *4.4. Limitations*

There are a few issues that should be addressed in the future. First, although the sample size of DOC patients in our study was very large for a single center, we only had a small number of the patients for the study predicting outcomes at 3 months post-scan. Additional studies with more patients will be needed to validate and confirm our findings. Second, the DTI technique is incapable of quantifying complex microstructural changes in the voxels with crossing fibers or partial volume effects and therefore may provide false-negative or false-positive results. To overcome these issues, new diffusion MRI techniques (e.g., diffusion spectrum imaging (Wedeen et al., 2008)) can be applied in the future. Finally, in the current study, we considered WM regions separately and found a widespread distribution of consciousness-related WM regions, suggesting that the level of consciousness reflects the integration of activity in multiple brain regions. Therefore, it would also very interesting to explore how the global organizational architecture of the WM network is related to the loss of consciousness in DOC patients in further investigations.

## 5. Acknowledgements

This work was supported by the National Science Foundation for Distinguished Young Scholars of China (grant number 81025013), China's National Strategic Basic Research Program ("973") grant (grant numbers 2012CB720700, 2010CB945500, 2012CB966300, and 2009CB941100), the National Natural Science Foundation of China (grant numbers 81322021 and 81571025), the Beijing Nova Program (grant number Z121110002512032), the Project for National 985 Engineering of China (grant number 985III-YFX0102), the "Dawn Tracking" Program of Shanghai Education Commission (grant number 10GG01), the Shanghai Natural Science Foundation (grant numbers: 08411952000 and 10ZR1405400), the National Natural Science Young Foundation in China (grant number: 81201033), the Shanghai Health Bureau (20114358), the 863 National Science and Technology Program (grant number: 2015AA020501), the Program for New Century Excellent Talents in University (NCET-10-0356) and the National Program for the Support of Top-Notch Young Professionals. Dr. Georg Northoff is supported by the Michael Smith Foundation, the CRC, and the CIHR. Jiaying Zhang is supported by the China Scholarship Council.

**Figure Legends**

**Fig. 1.** Fourteen consciousness-related WM ROIs detected with FA ANCOVA. The mean FA differed across the levels of consciousness according to the group comparisons (corrected $p < 0.05$, Bonferroni correction). In each subfigure, the left column shows a WM ROI in standard space, with the corresponding brain areas indicated in red, and the graphs on the right show the fitted mean and standard deviation of the FA of the corresponding WM ROI for the four DOC subgroups. Blue represents COMA; red, UWS/VS; green, MCS; and purple, PC. * indicates $p < 0.05$, and ** indicates $p < 0.01$.

**Fig. 2.** Correlations between the fitted mean FA and the clinical measure of consciousness level, GCS score, for each of the 14 WM ROIs. The light blue dots represent the fitted mean FA of all of the DOC patients, and the dark blue dots represent the fitted mean FA of the DOC patients after the exclusion of those with a GCS score of 15.

**Fig. 3.** Correlations between the fitted mean FA and the clinical measure of consciousness level, CRS-R score, for each of the 14 WM ROIs. The light blue dots represent the fitted mean FA of all of the DOC patients, and the dark blue dots represent the mean fitted FA of the DOC patients after the exclusion of those with a CRS-R score of 23.

**Fig. 4.** DTI parameters predicted recovery outcomes at 3 months after MRI scan. (a) Fitted mean FA in WM ROI SS.R and (c) fitted mean RD in WM ROI SCP.L for the two groups with different functional outcomes – regaining consciousness or not. The blue dots (GOS score < 3) represent the fitted diffusion parameters of the better outcome group, and the red dots (GOS score ≥ 3) represent the fitted diffusion parameters of the other group. The fitted mean FA values in the two groups differed significantly (uncorrected, $p < 0.05$). The initial clinical scores (GCS and CRS-R scores) were separately included in the statistical model to reduce the effect of the initial level of consciousness. In addition, (b) and (d) show the ROC curves for the fitted mean FA of SS.R and the fitted mean RD of SCP.L, respectively.

**Tables**

**Table 1. Demographic and clinical characteristics of all DOC patients**

| Diagnostic categories[NO.] | COMA[8(3)] | UWS/VS[8(1)] | MCS[14(3)] | PC[25(2)] | Statistic | p |
|---|---|---|---|---|---|---|
| Age: Mean±Std | 42.3±8.7 | 46.4±13.9 | 43.1±16.7 | 38.2±14.5 | $F_{3,51}=0.79$ | 0.506 |
| Gender: male/female | 6/2 | 3/5 | 11/3 | 18/7 | $\chi^2=4.56$ | 0.207 |
| Etiology: TBI/non-TBI | 8/0 | 6/2 | 13/1 | 22/3 | $\chi^2=2.85$ | 0.416 |
| Days post-ictus: Median (range) | 20 (13-42) | 96 (10-182) | 20.5 (15-98) | 58 (4-167) | =4.33 | 0.009* |

There were no differences among the four DOC groups in age at scan, gender, or etiology distribution. The days post-ictus differed among the four groups. The bracketed number indicates the number of patients who had a second qualified MRI scan. The etiology of each non-TBI patient was either ICH or IHI.

**Table 2. Demographic and clinical characteristics of the DOC patients with GOS < 3 and GOS ≥ 3**

| Items(NO.) | GOS < 3(8) | GOS ≥ 3(8) | p |
| --- | --- | --- | --- |
| Age: Mean±Std | 49.8±10.1 | 38.9±10.4 | 0.052 |
| Gender: male/female | 4/4 | 5/3 | 0.626 |
| Etiology: TBI/non-TBI | 6/2 | 8/0 | 0.143 |
| Days post-ictus: Median (range) | 43 (10-182) | 23 (13-168) | 0.812 |

There were no differences between the two groups in gender, etiology distribution, or days post-ictus. Age differed between the groups. The etiology of the non-TBI patients was either ICH or IHI.

**Table 3.** WM deficits in DOC using ANCOVA

| DTI Parameter | WM Region of Interest | $F_{3,48}$ | p |
|---|---|---|---|
| FA | middle cerebellar peduncle | 6.348 | 0.001 |
| | right corticospinal tract | 6.549 | 0.0008 |
| | right superior cerebellar peduncle | 7.640 | 0.0003 |
| | left superior cerebellar peduncle | 7.626 | 0.0003 |
| | right cerebral peduncle | 8.413 | 0.0001 |
| | body of corpus callosum | 8.675 | 0.0001 |
| | splenium of corpus callosum | 7.332 | 0.0004 |
| | left cingulum | 7.051 | 0.0005 |
| | fornix (column and body) | 6.642 | 0.0008 |
| | left stria terminalis | 6.569 | 0.0008 |
| | right internal capsule (retrolenticular part) | 6.543 | 0.0008 |
| | right superior fronto-occipital fasciculus | 6.893 | 0.0006 |
| | right sagittal stratum | 7.885 | 0.0002 |
| | left uncinate fasciculus | 6.834 | 0.0006 |
| AD | right cerebral peduncle | 2.830 | 0.048 |
| | splenium of corpus callosum | 2.976 | 0.041 |
| | left stria terminalis | 3.888 | 0.014 |
| RD | right superior cerebellar peduncle | 5.270 | 0.003 |
| | left superior cerebellar peduncle | 4.435 | 0.008 |
| | right cerebral peduncle | 4.041 | 0.012 |
| | body of corpus callosum | 5.305 | 0.003 |
| | splenium of corpus callosum | 5.512 | 0.002 |
| | left cingulum | 5.287 | 0.003 |
| | fornix (column and body) | 3.529 | 0.022 |
| | left stria terminalis | 5.136 | 0.004 |
| | right internal capsule (retrolenticular part) | 4.386 | 0.008 |
| | right sagittal stratum | 6.910 | 0.001 |

| | left uncinate fasciculus | *5.003* | *0.004* |

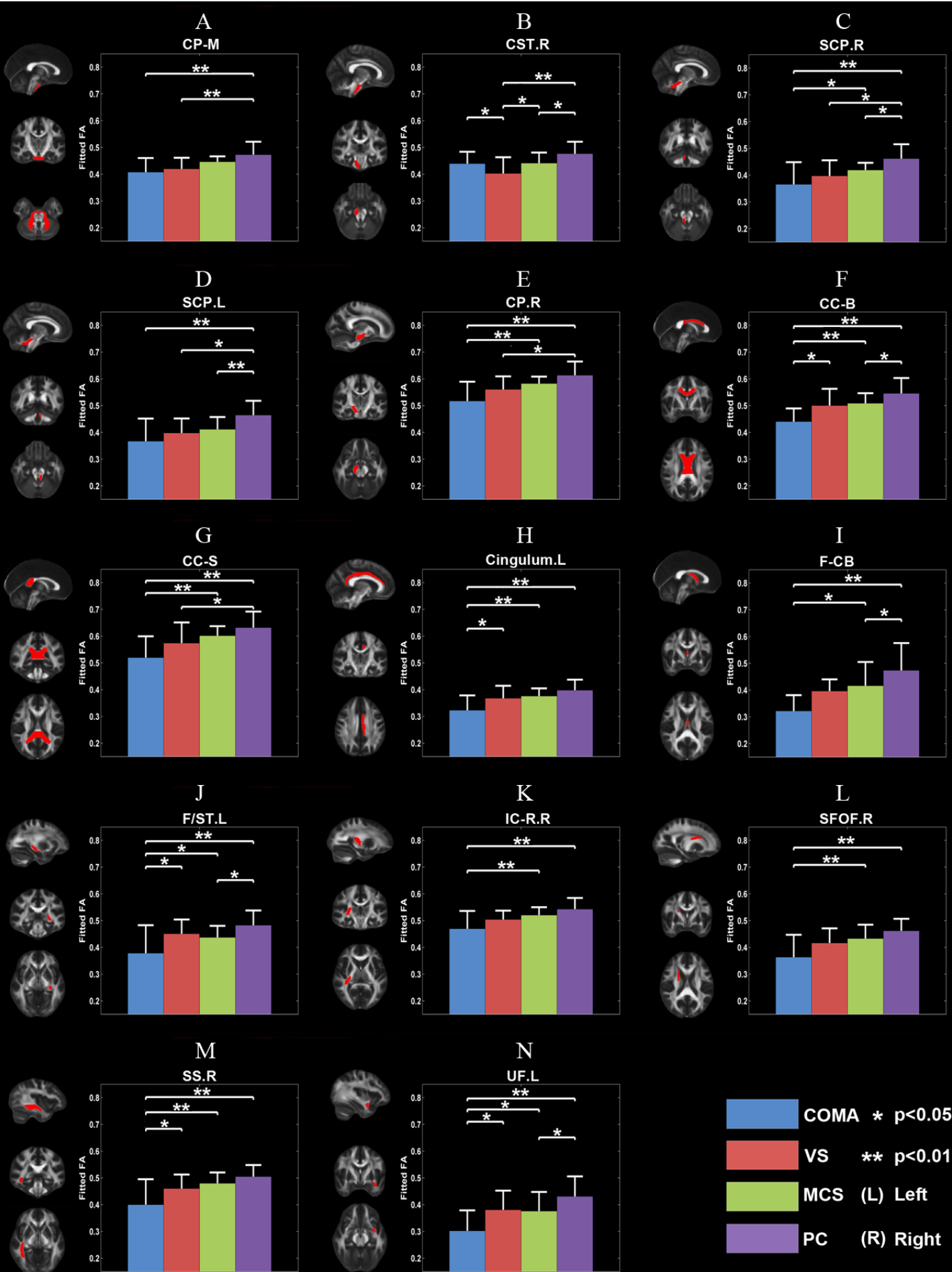

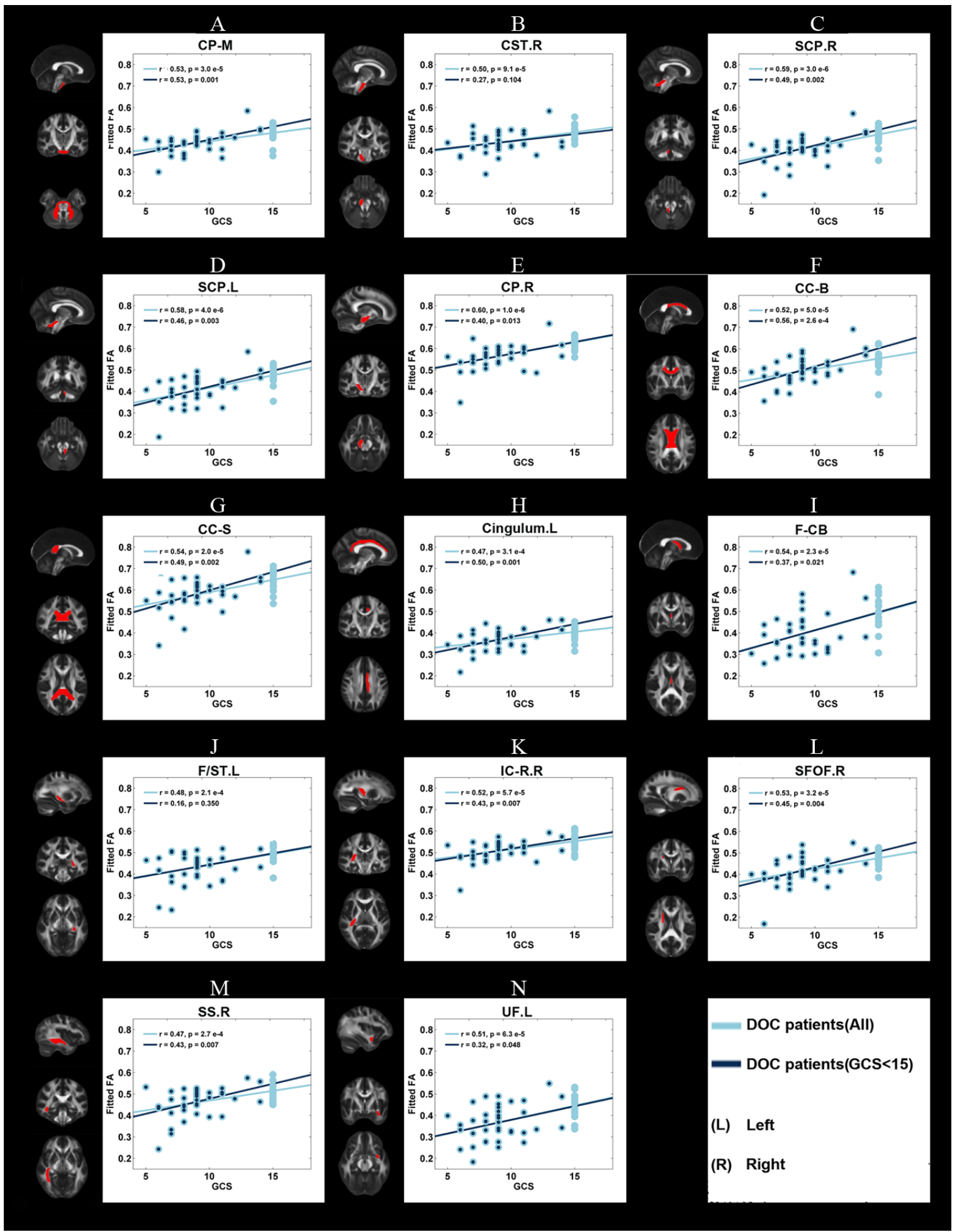

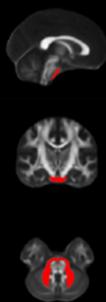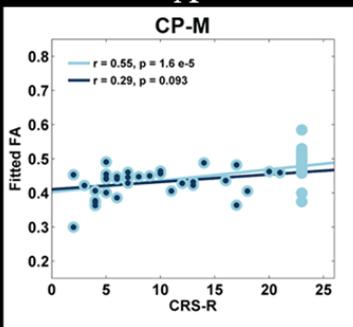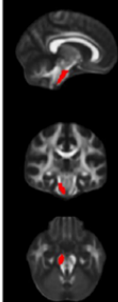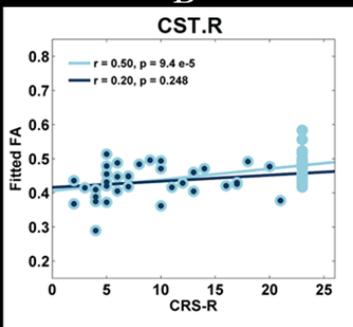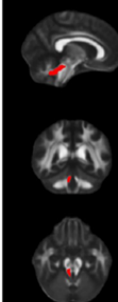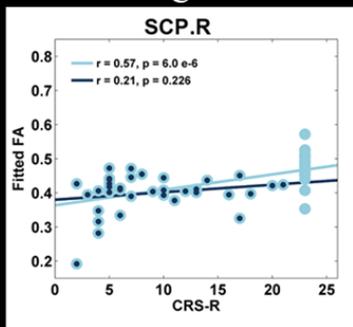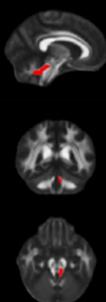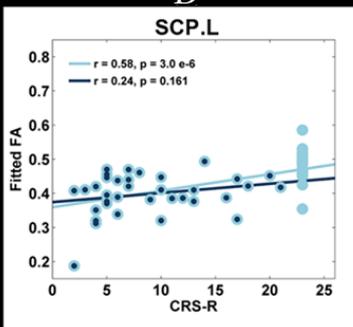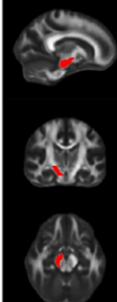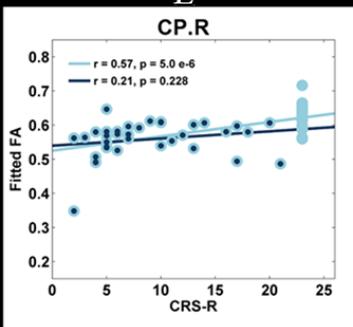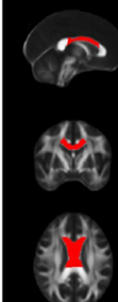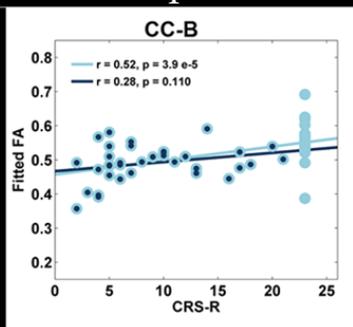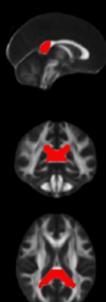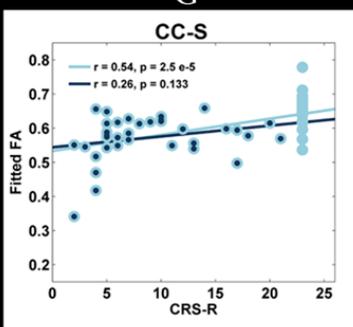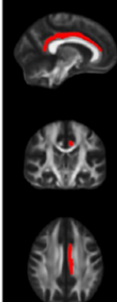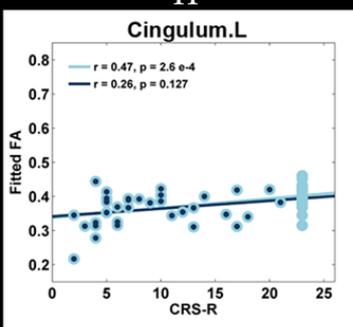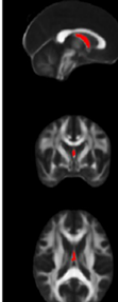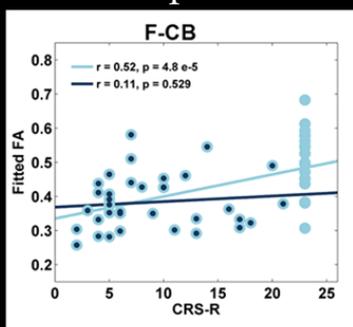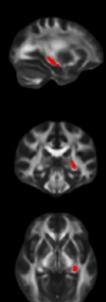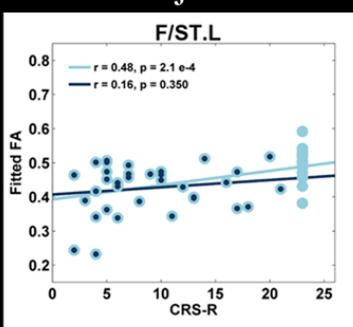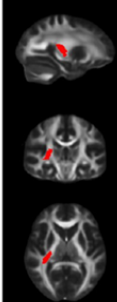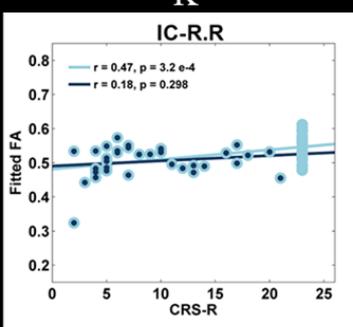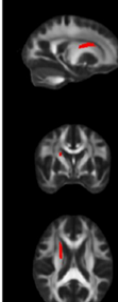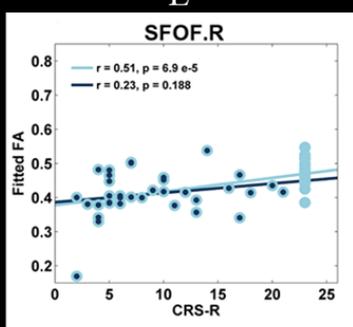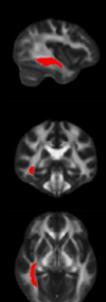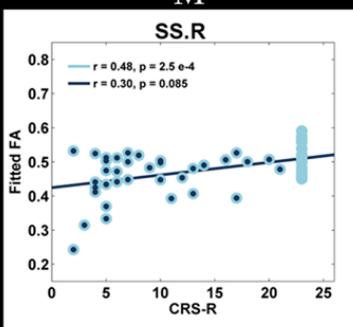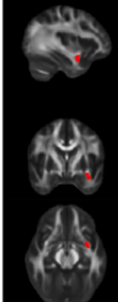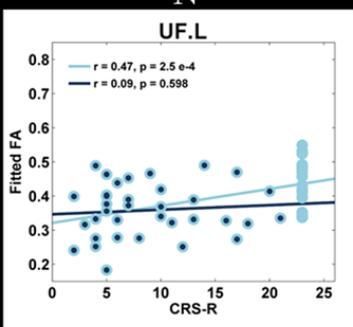

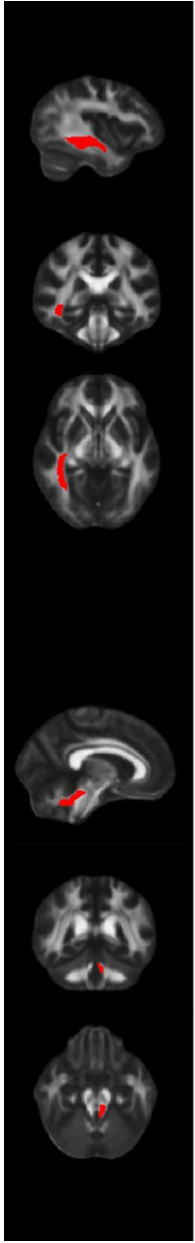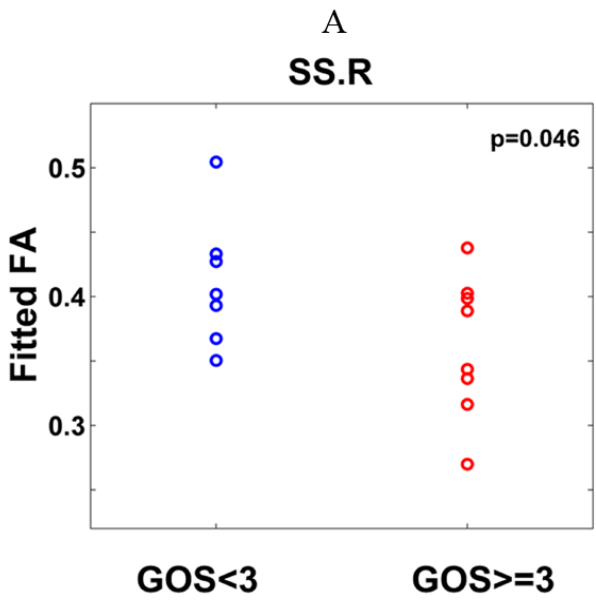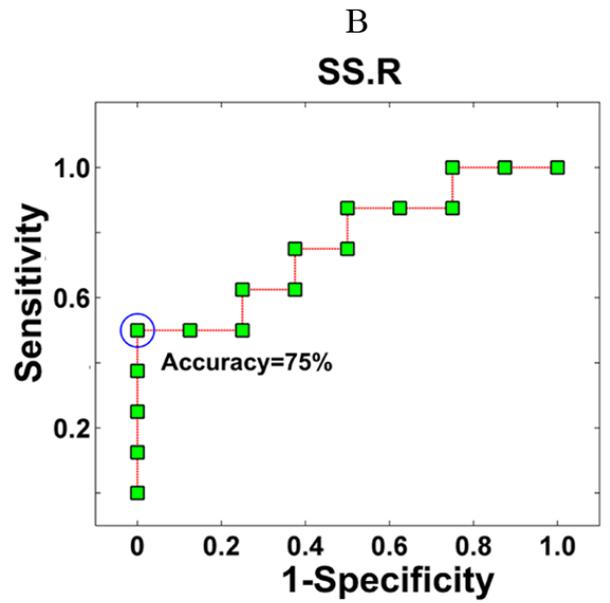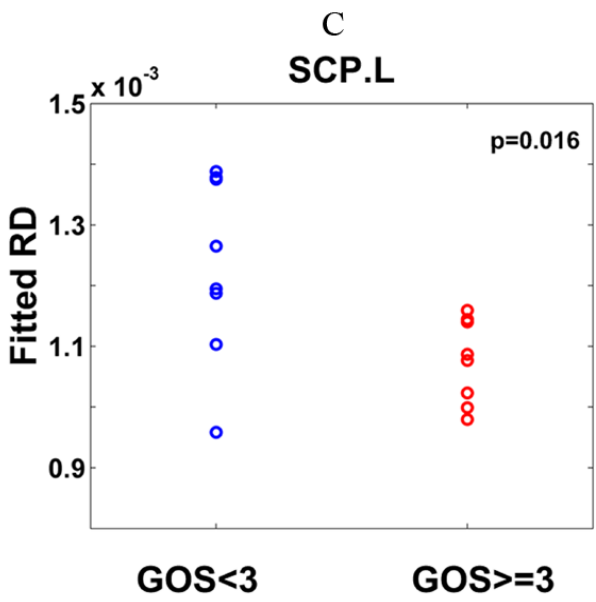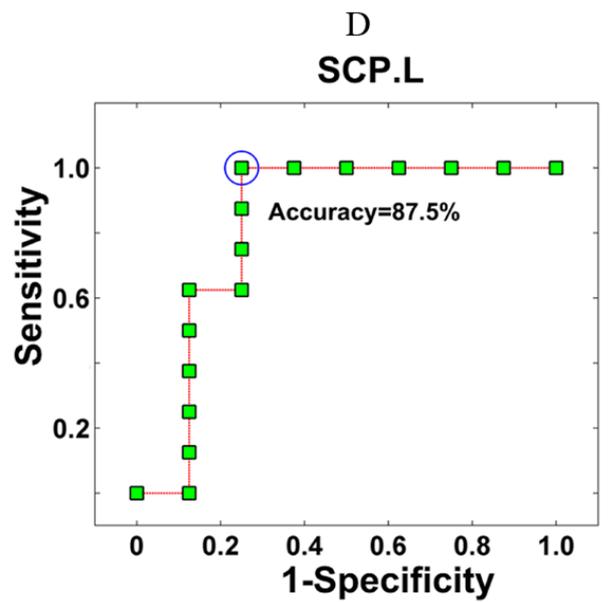

# Supplementary Material

**Table 1. Consciousness-related WM ROIs defined in the JHU atlas and their corresponding abbreviations**

| WM labels in the JHU atlas | Abbreviations |
| --- | --- |
| Middle Cerebellar Peduncle | CP-M |
| Body of Corpus Callosum | CC-B |
| Splenium of Corpus Callosum | CC-S |
| Fornix (Column and Body part) | F-CB |
| Right Corticospinal Tract | CST.R |
| Right Superior Cerebellar Peduncle | SCP.R |
| Left Superior Cerebellar Peduncle | SCP.L |
| Right Cerebral Peduncle | CP.R |
| Right Retrolenticular Part of Internal Capsule | IC-R.R |
| Right Sagittal Stratum | SS.R |
| Left Cingulum | Cingulum.L |
| Left Stria Terminalis | F/ST.L |
| Right Superior Fronto-Occipital Fasciculus | SFOF.R |
| Left Uncinate Fasciculus | UF.L |

**Table 2. The detailed clinical information of the participants in our study**

| Patient | Age | Gender | Handness | Etiology | Days post ictus | GCS | CRS-R | GOS |
|---|---|---|---|---|---|---|---|---|
| COMA1 | 46 | male | Right | TBI | 19 | 5 | 2 | 2 |
| COMA2 | 24 | male | Right | TBI | 15 | 8 | 7 | 5 |
| COMA3 | 50 | male | Right | TBI | 42 | 7 | 4 | 2 |
| COMA4 | 37 | male | Right | TBI | 21 | 7 | 3 | 3 |
| COMA5 | 46 | male | Right | TBI | 25 | 6 | 2 | 4 |
| COMA6 | 47 | male | Right | TBI | 13 | 7 | 5 | 3 |
| COMA7 | 39 | female | Right | TBI | 27 | 8 | 6 | 2 |
| COMA8 | 49 | female | Right | TBI | 17 | 7 | 5 | 3 |
| UWS/VS1 | 49 | male | Right | TBI | 140 | 6 | 5 | 2 |
| UWS/VS2 | 36 | female | Right | non-TBI | 182 | 9 | 5 | 2 |
| UWS/VS3 | 59 | male | Right | TBI | 44 | 8 | 4 | 2 |
| UWS/VS4 | 67 | female | Right | TBI | 10 | 6 | 4 | 2 |
| UWS/VS5 | 38 | female | Right | TBI | 31 | 9 | 6 | 3 |
| UWS/VS6 | 23 | female | Right | TBI | 168 | 8 | 4 | 3 |
| UWS/VS7 | 52 | female | Right | non-TBI | 52 | 7 | 5 | 2 |
| UWS/VS8 | 47 | male | Right | TBI | 163 | 8 | 5 | 3 |
| MCS1 | 30 | female | Right | TBI | 26 | 9 | 10 | |
| MCS2 | 24 | male | Right | TBI | 16 | 9 | 12 | |
| MCS3 | 18 | male | Right | TBI | 31 | 9 | 6 | |
| MCS4 | 49 | male | Right | TBI | 98 | 10 | 16 | |
| MCS5 | 55 | male | Right | TBI | 18 | 9 | 7 | |
| MCS6 | 46 | male | Right | TBI | 15 | 9 | 7 | |
| MCS7 | 63 | female | Right | TBI | 28 | 9 | 13 | |
| MCS8 | 60 | male | Right | TBI | 20 | 9 | 10 | |
| MCS9 | 18 | male | Right | TBI | 15 | 9 | 14 | |
| MCS10 | 31 | female | Right | TBI | 73 | 10 | 9 | |
| MCS11 | 67 | male | Right | non-TBI | 18 | 9 | 8 | |
| MCS12 | 37 | male | Right | TBI | 82 | 8 | 13 | |
| MCS13 | 48 | male | Right | TBI | 19 | 9 | 10 | |
| MCS14 | 57 | male | Right | TBI | 21 | 10 | 11 | |
| PC1 | 42 | male | Right | TBI | 111 | 11 | 18 | |
| PC2 | 16 | female | Right | TBI | 16 | 15 | 23 | |
| PC3 | 47 | male | Right | TBI | 98 | 15 | 23 | |
| PC4 | 17 | male | Right | TBI | 28 | 15 | 23 | |
| PC5 | 70 | female | Right | TBI | 4 | 15 | 23 | |
| PC6 | 50 | male | Right | TBI | 20 | 15 | 23 | |
| PC7 | 58 | male | Right | TBI | 10 | 11 | 20 | |
| PC8 | 52 | female | Right | TBI | 127 | 15 | 23 | |
| PC9 | 32 | male | Right | TBI | 68 | 14 | 23 | |
| PC10 | 41 | male | Right | non-TBI | 63 | 15 | 23 | |
| PC11 | 52 | male | Right | TBI | 6 | 15 | 23 | |
| PC12 | 29 | female | Right | TBI | 118 | 14 | 23 | |
| PC13 | 30 | male | Right | TBI | 19 | 15 | 23 | |
| PC14 | 28 | female | Right | TBI | 13 | 15 | 23 | |
| PC15 | 22 | male | Right | TBI | 66 | 15 | 23 | |
| PC16 | 64 | male | Right | TBI | 7 | 15 | 23 | |
| PC17 | 28 | female | Right | TBI | 4 | 15 | 23 | |
| PC18 | 42 | male | Right | non-TBI | 5 | 11 | 17 | |
| PC19 | 49 | male | Right | TBI | 167 | 13 | 23 | |
| PC20 | 29 | female | Right | TBI | 5 | 15 | 23 | |
| PC21 | 22 | male | Right | non-TBI | 166 | 15 | 23 | |
| PC22 | 46 | male | Right | TBI | 58 | 15 | 23 | |
| PC23 | 25 | male | Right | TBI | 143 | 15 | 23 | |
| PC24 | 32 | male | Right | TBI | 86 | 12 | 21 | |
| PC25 | 33 | male | Right | TBI | 119 | 11 | 17 | |

**Figure Legends for Supplementary Figures.**

**Fig. 1. Clinical scores of consciousness level in the DOC patient population.** (A) The distribution of GCS in the DOC patient population. (B) The distribution of CRS-R in the DOC patient population. (C) The means and standard deviations of GCS and CRS-R for each DOC subgroup; blue represents COMA, red VS, green MCS, and purple PC.

**Fig. 2. Three of 14 consciousness-related WM ROIs detected with further AD ANCOVAs.** These ROIs were significantly different in mean AD across levels of consciousness by group comparisons ($p < 0.05$, uncorrected). For each subfigure, the left column shows a WM ROI in the standard space indicated by brain areas in red, and the right part shows the fitted mean and standard deviation of AD in the corresponding WM ROI for four DOC subgroups. Blue represents COMA, red VS, green MCS, and purple PC. * indicates that $p < 0.05$, and ** indicates that $p < 0.01$.

**Fig. 3. Eleven of 14 consciousness-related WM ROIs detected with RD ANCOVAs.** These ROIs were significantly different in mean RD across levels of consciousness by group comparisons ($p < 0.05$, uncorrected). For each subfigure, the left column shows a WM ROI in the standard space indicated by brain areas in red, and the right part shows the fitted mean and standard deviation of RD in the corresponding WM ROI for four DOC subgroups. Blue represents COMA, red VS, green MCS, and purple PC. * indicates that $p < 0.05$, and ** indicates that $p < 0.01$.

**Fig. 4. Correlations between fitted mean AD and the clinical measure of the consciousness level – GCS for each of the three WM ROIs.** The light blue represents the fitted mean AD of all the DOC patients, and the dark blue represents the fitted mean AD of the DOC patients after the exclusion of those with a GCS score of 15.

**Fig. 5. Correlations between fitted mean AD and the clinical measure of the consciousness level – CRS-R for each of the three WM ROIs.** The light blue represents the fitted mean AD of all the DOC patients, and the dark blue represents

the fitted mean AD of the DOC patients after the exclusion of those with a CRS-R score of 23.

**Fig. 6. Correlations between fitted mean RD and the clinical measure of the consciousness level – GCS for each of the 11 WM ROIs.** The light blue represents the fitted mean RD of all the DOC patients, and the dark blue represents the fitted mean RD of the DOC patients after the exclusion of those with a GCS score of 15.

**Fig. 7. Correlations between fitted mean RD and the clinical measure of the consciousness level – CRS-R for each of the 11 WM ROIs.** The light blue represents the fitted mean RD of all the DOC patients, and the dark blue represents the fitted mean RD of the DOC patients after the exclusion of those with a CRS-R score of 23.

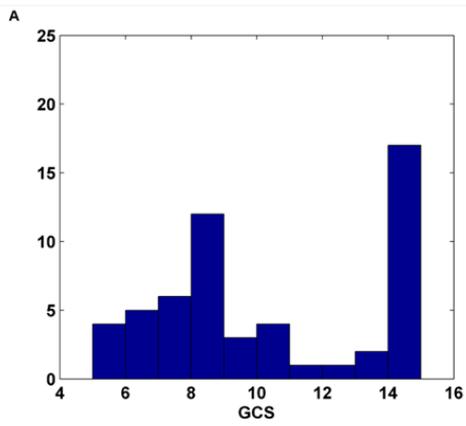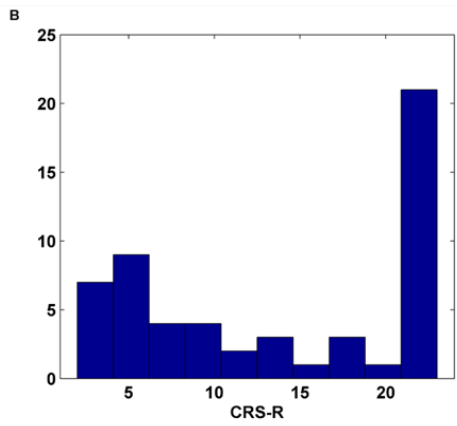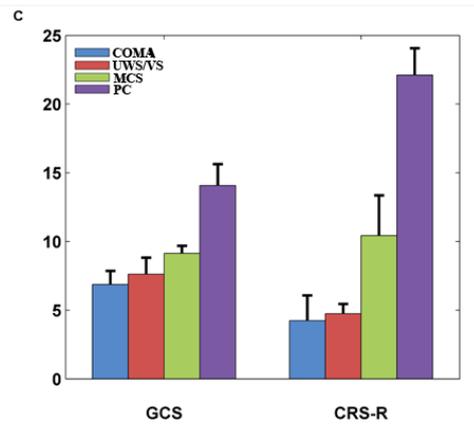

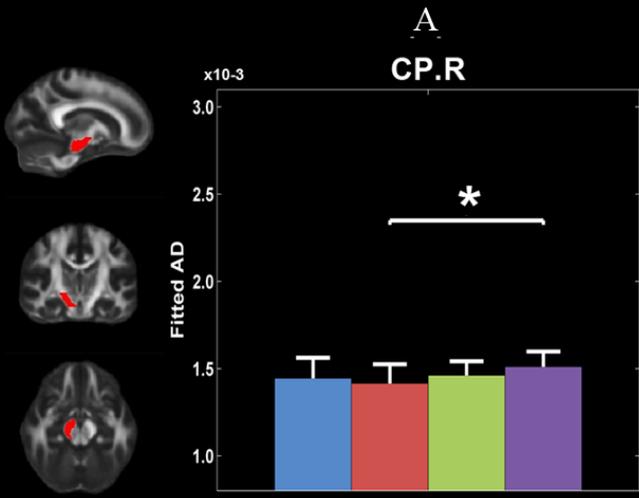
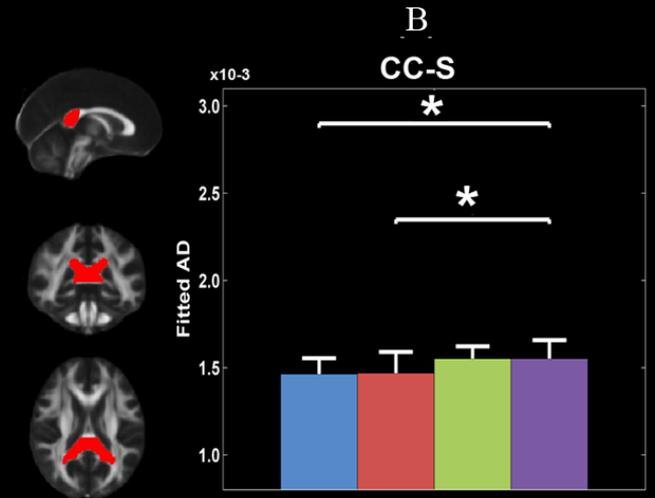
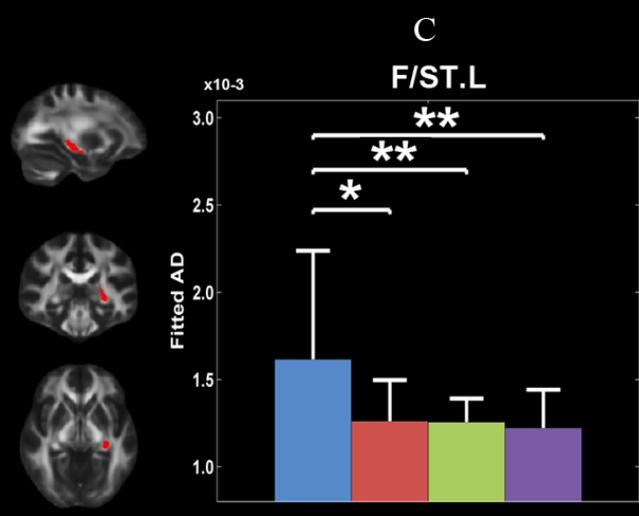

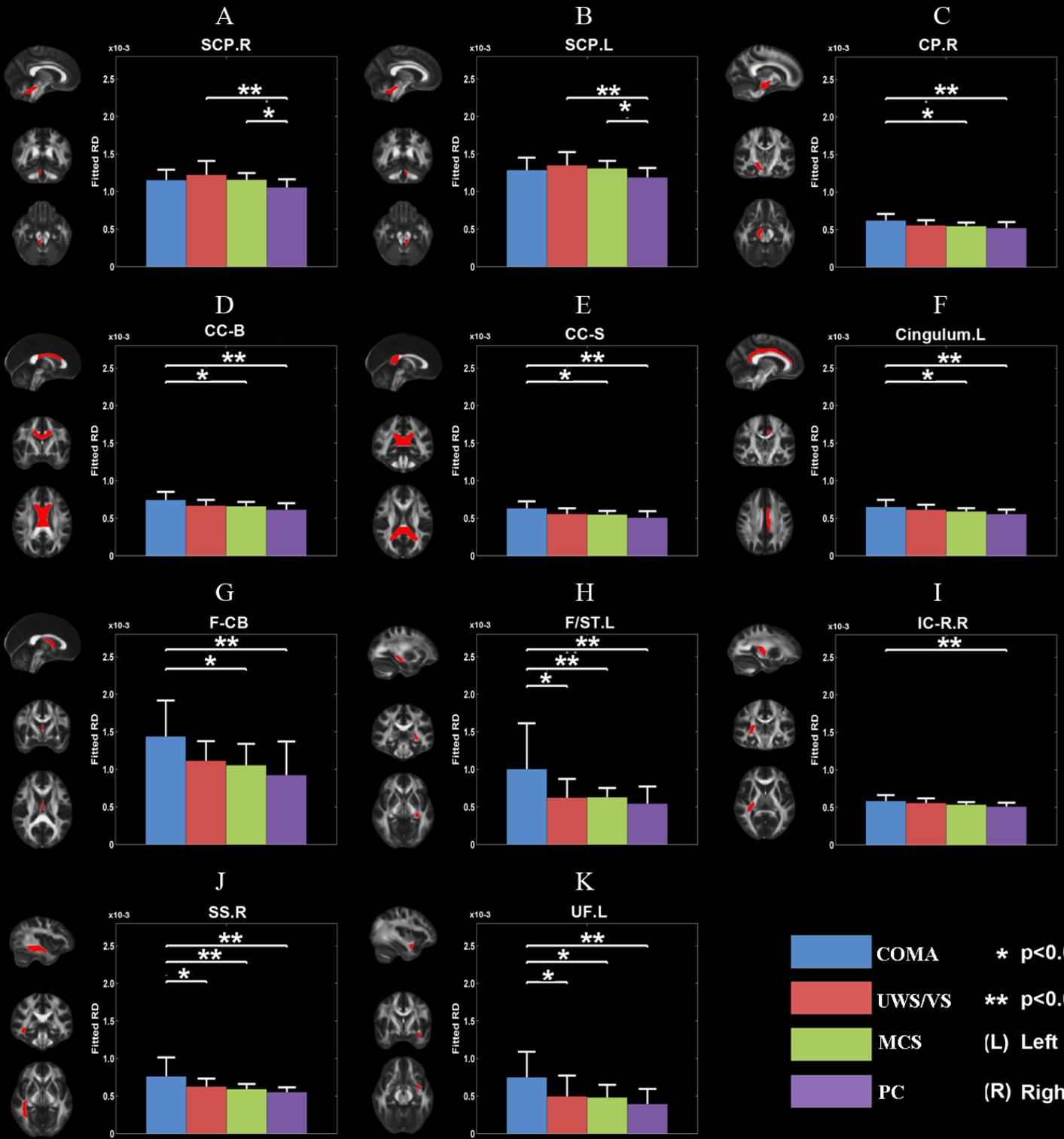

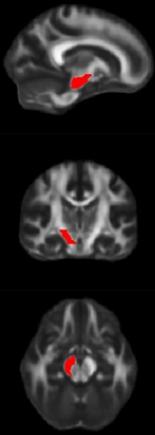 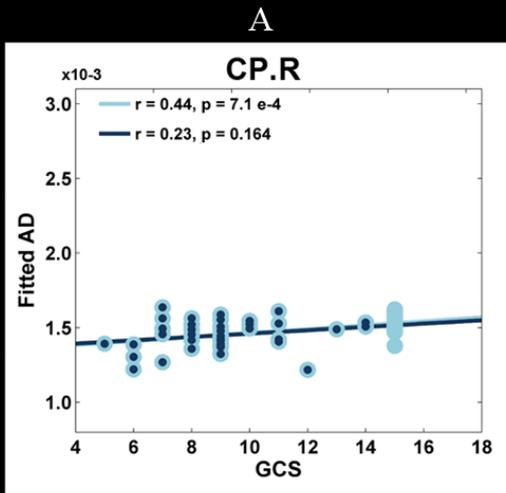 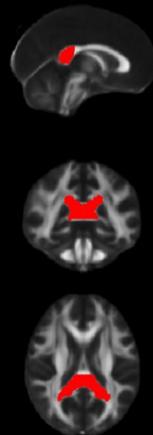 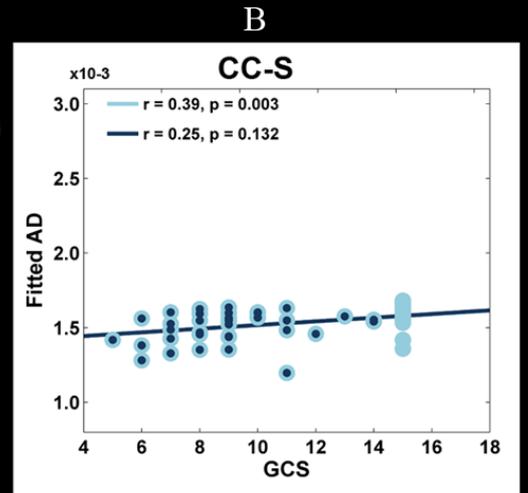
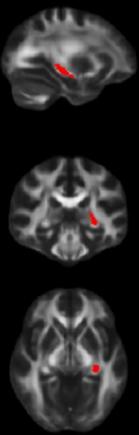 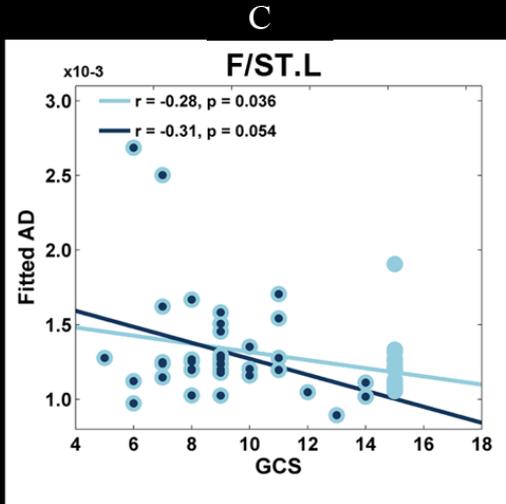

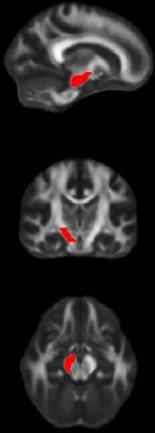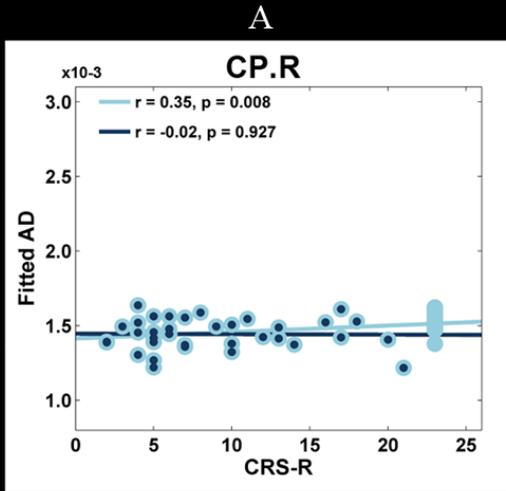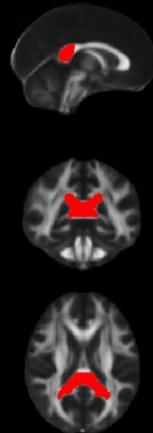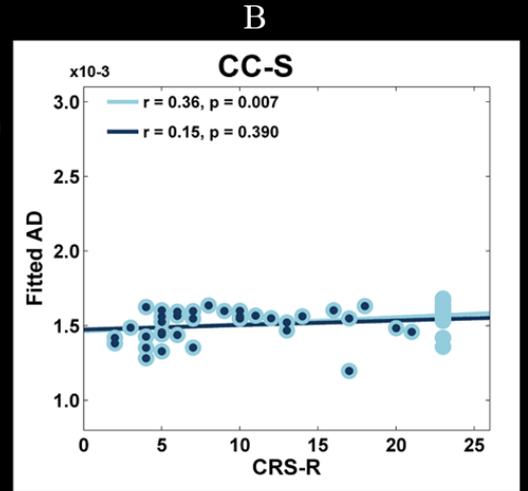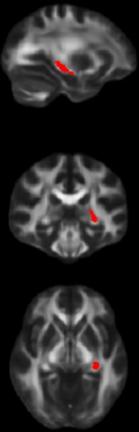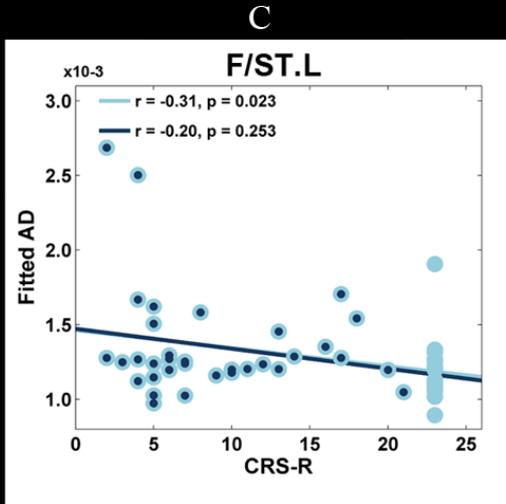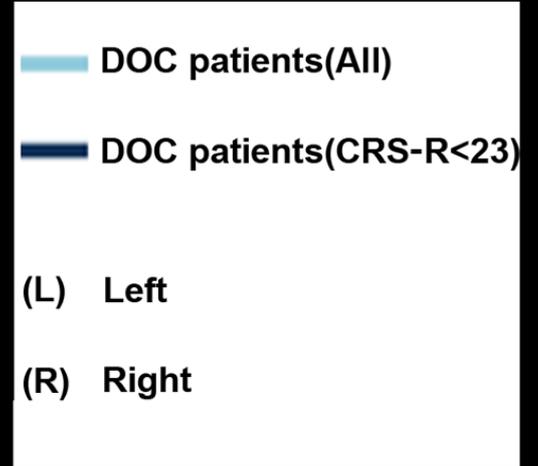

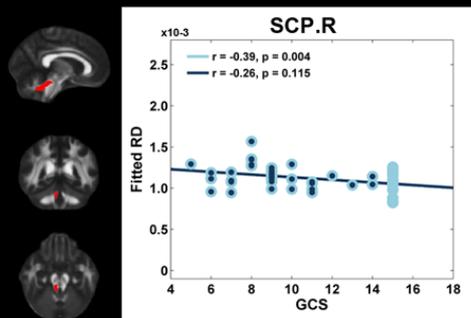
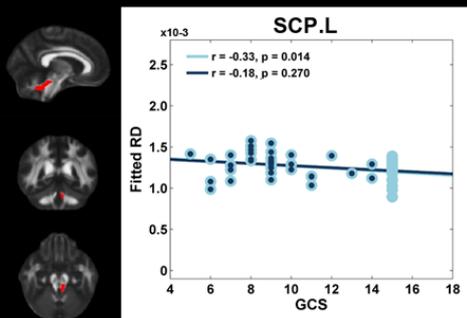
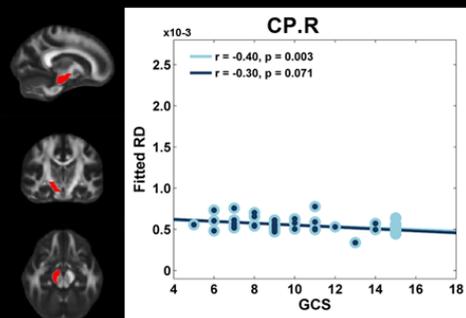
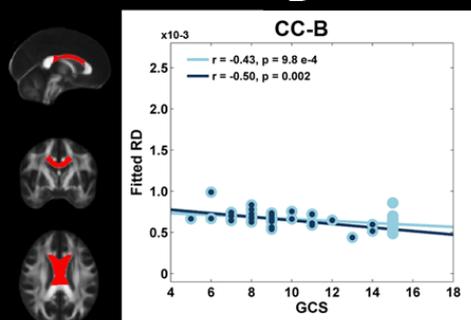
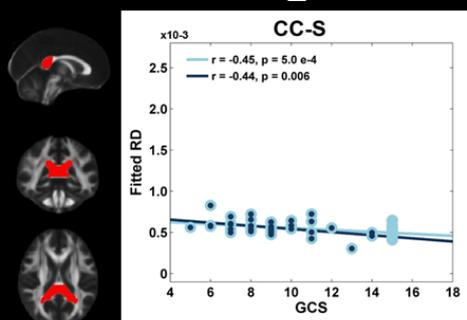
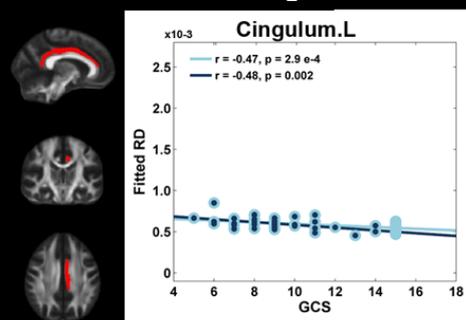
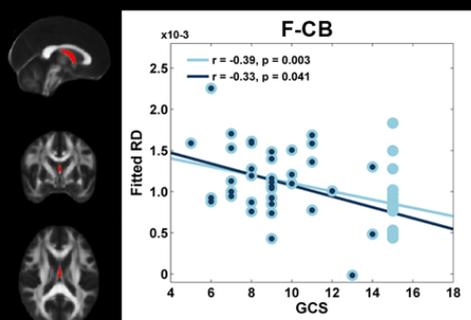
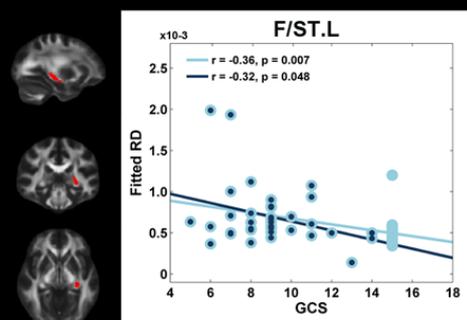
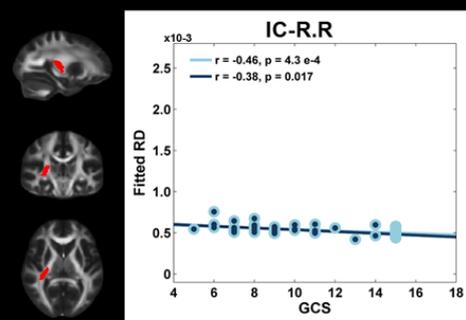
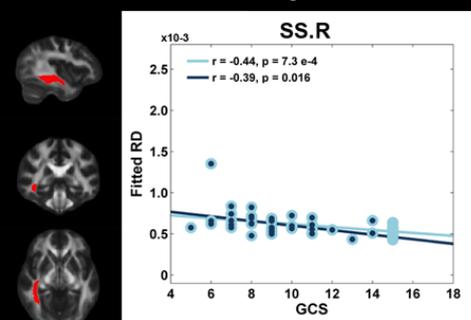
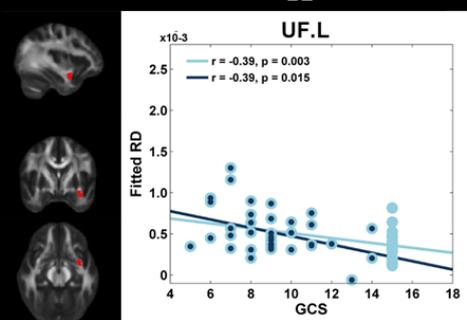

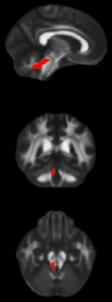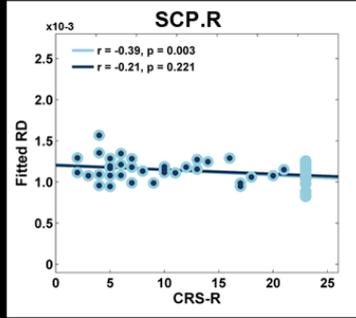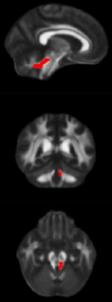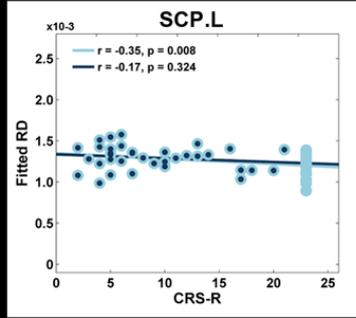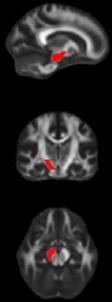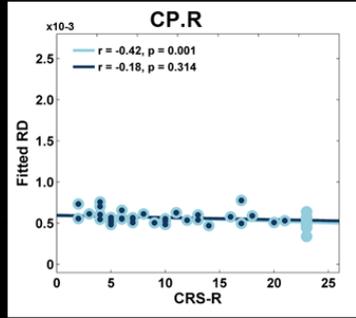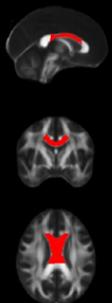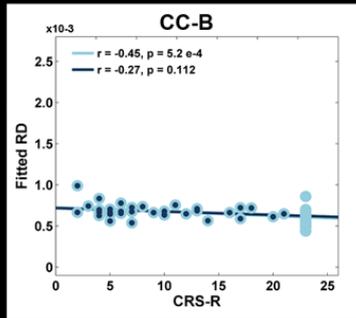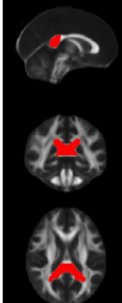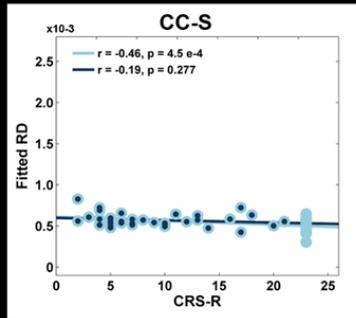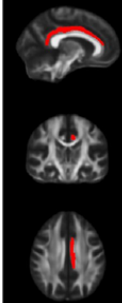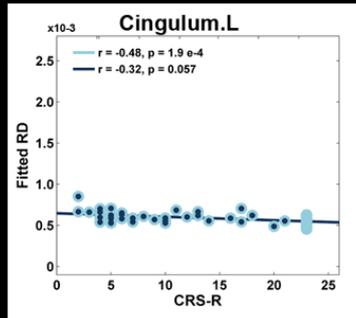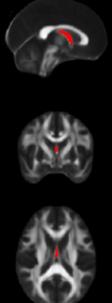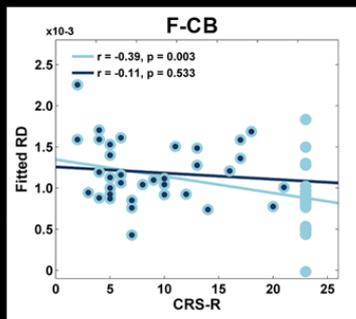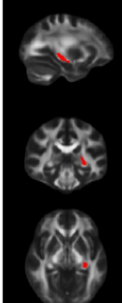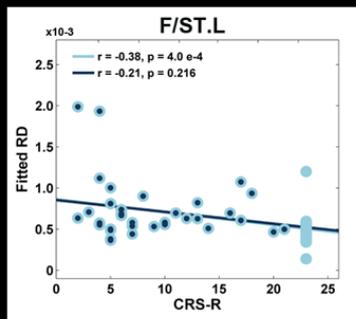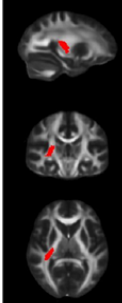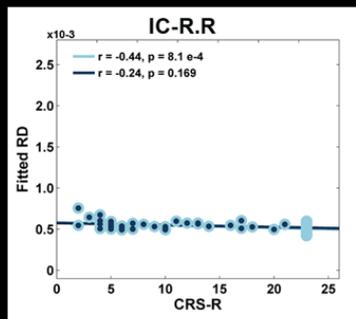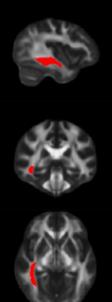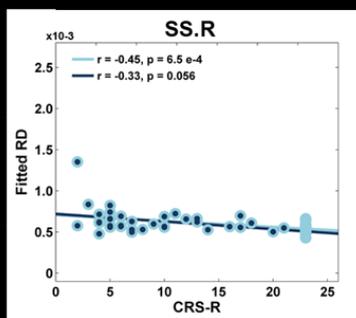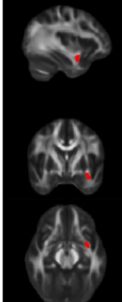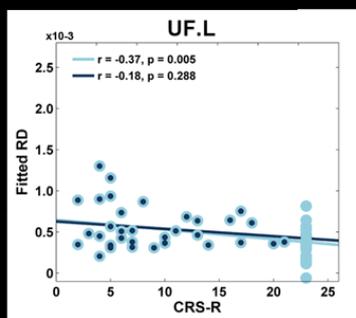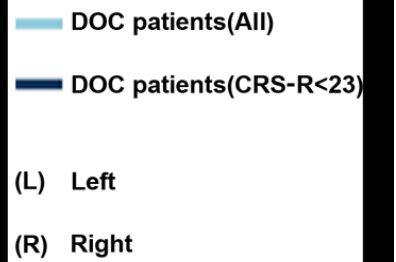